\documentclass[11pt,fleqn]{article}
\usepackage{multicol}
\usepackage{graphicx}
\newcommand\feature[1]{\vskip2.7mm
\noindent\underline{\em #1}\vskip2.3mm\nobreak}

\begin{document}
\sf

\centerline{\Huge $U_{\! e 3}$ from physics above the GUT scale}

\vspace{7mm}

\centerline{\large
Francesco Vissani,
Mohan Narayan, 
Veniamin Berezinsky
}
\centerline{\em INFN, Laboratori Nazionali del Gran Sasso,
I-67010 Assergi (AQ), Italia}
\vspace{8mm}

\centerline{\large\sc Abstract}
\begin{quote}
\small
We consider non-renormalizable $1/M_X$ 
interaction terms as a perturbation of the 
conventional neutrino mass matrix. 
Particular attention is given to the gravitational 
interaction with $M_X=M_{\rm Pl}$. 
We find that for the degenerate
neutrino mass spectrum, the considered perturbation generates a
non-zero $U_{e 3}$ which is within reach of the high 
performance neutrino factories and just on the
borderline to be of interest for supernova physics. For the hierarchical
mass spectrum this effect is small. For $1/M_X$ interaction terms with 
$M_X$ about the GUT scale, 
a detectable $U_{e3}$ term is induced
for the hierarchical mass spectra also. 
Numerical estimates are given for all the above mentioned cases 
and renormalization effects are considered.

\end{quote}
\rm 

\section*{Introduction}
One of most important issues in neutrino physics is the
magnitude of $U_{e 3}$ or equivalently of the mixing angle $\theta_{13}$. 
The actual value of this parameter is of great interest to various
aspects of neutrino physics:
To theory, 
to the search for CP violation in the next generation  
of long baseline oscillation experiments, 
to interpret supernova neutrino signals, {\em etc}.
The only solid information we have on this 
parameter is the upper bound that is 
based on the CHOOZ experiment \cite{CHOOZ}, 
that is $\theta_{13} \leq 7^\circ$  at 1 
sigma ($13^\circ$ at 3 sigma). 

In this paper, we point out that a tentative lower bound on 
this parameter arises from physics above the grand unified 
theory (GUT) scale, in particular from the
gravitational interaction of neutrinos. 
Let us describe this idea.
Most probably, grand unified dynamics generates the main part
of the neutrino mass matrix. However, contributions from other sources
are likely to exist. Specifically,
gravitational interactions 
\cite{planck1,planck2,planck4} can produce 
additional terms in the mass of neutrinos,
and these can affect the size of $\theta_{13}$.

The relevant gravitational dimension-5 operator for the spinor 
$SU(2)_L$ isodoublets,\footnote{Here and everywhere below we 
use Greek letters $\alpha,\beta,$... for the flavor states 
and Latin letters $i,j,k$... for the mass states.} 
$\psi_\alpha=(\nu_\alpha,\ell_\alpha)$ 
and the scalar one, $\varphi=(\varphi^+,\varphi^0)$, 
can be written with the operators 
introduced by Weinberg \cite{d=5} as
\begin{equation}
{\cal L_{\rm grav}} = \frac{\lambda}{M_{\rm Pl}}
({\psi}_{Aa\alpha}\: \epsilon_{AC}\: \varphi_C)\: C_{ab}^{\scriptscriptstyle -1}\:
({\psi}_{Bb\beta}\: \epsilon_{BD}\: \varphi_D) +h.c.,
\label{grav}
\end{equation}
where $M_{\rm Pl}= 1.2\times 10^{19}$~GeV  is the Planck mass,
and $\lambda$ is a number ${\cal O}(1)$.
In eq.(\ref{grav}),
all indices are explicitly shown:
the Lorentz indices $a,b=1,2,3,4$ 
are contracted with the charge conjugation matrix $C$,
the $SU(2)_L$ isospin indices $A,B,C,D=1,2$ 
are contracted with $\epsilon=i\sigma_2$; 
$\sigma_m$ ($m=1,2,3$) are the Pauli matrices.
After spontaneous electroweak
symmetry breaking, the Lagrangian~(\ref{grav})
generates additional terms of neutrino mass:
${\cal L_{\rm mass}}= \lambda\: {v^2}/{M_{\rm Pl}}\ \nu_{\alpha}
C^{\scriptscriptstyle -1}\nu_{\beta}$,
where $v$=174~GeV is the vev of electroweak 
symmetry breaking.  We assume that the
gravitational interaction is ``flavour blind'', 
{\em i.e.}~$\lambda$ does not 
contain $\alpha,\beta$ indices. 
In tis case, the contribution to the neutrino 
mass matrix is of the order of:
\begin{equation}
\mu \textstyle
\left( \begin{array}{lcr}
  1 & \!1\! & 1 \\[-.5ex]
  1 & \!1\! & 1 \\[-.5ex]
  1 & \!1\! & 1 
\end{array} \right), 
\label{textm}
\end{equation}
where the scale $\mu$ (`scale of perturbation') is 
\begin{equation}
\mu=v^2/M_{\rm Pl}= 2.5 \times 10^{-6}~\mbox{eV}. 
\label{mu}
\end{equation}
In our calculations, we take 
eq.(\ref{textm}) as a `contribution of perturbation' to the 
main part of neutrino mass matrix, that is generated by GUT.
In other words, our results are normalized to 
the case $\lambda=1/2$; more discussion on $\lambda$ 
follows.

However, there might be some other interactions of the form (\ref{grav})
with a scale $M_X$ less than $M_{\rm Pl}$, {\em e.g.}\ 
from string 
compactification, non-perturbative dynamics, 
flavor physics or simply higher 
GUT scales. These interactions are not necessarily 
flavor blind. This implies not only a different scale of perturbation
$\mu= v^2/M_X$, but also a modified structure for
the matrix of perturbation, that we denote by 
$\lambda_{\alpha\beta}$ henceforth. In the case of quantum gravity,
$\lambda_{\alpha\beta}=\lambda$.

In this paper, we evaluate the effects of such  operators 
on the conventional neutrino mass matrix, given for example by the 
seesaw mechanism \cite{d=5,seesaw1,seesaw2,seesaw10}. 
We demonstrate that the gravitational perturbation 
generates a non-zero $U_{e 3}$ even if the 
unperturbed
mass  matrix has $U_{e 3}=0$. This non-zero value 
of $U_{e 3}$  can be considered 
as a lower bound imposed by gravitational effects 
and can be translated into a lower
bound for the angle $\theta_{13}$. 
The operators with the scale $M_X < M_{\rm Pl}$,
but higher than the GUT scale (string scales etc.) 
provides much stronger effects 
which are of potential interest to present 
and future observations.  For reference and for definiteness we 
recall that the supersymmetric 
unification scale is $M_{GUT}\sim 2\times 10^{16}$ GeV. However 
the considerations outlined below are not tied to any 
particular value of $M_{GUT}$.

Before passing to the calculations, we 
discuss in more detail the coefficient $\lambda$ in eq.(\ref{grav}).
The operator in eq.(\ref{grav}) is a term of the 
effective Lagrangian produced in quantum gravity, 
thus the coefficient $\lambda$ could be calculated
if the details of this theory were fixed;
but this is not the case at present. 
There is another way to see 
that the coefficient $\lambda$ is 
defined with considerable uncertainty. 
Indeed, an $SU(2)$ Fiertz  transformation yields 
$(\psi^t\: \sigma_2 \vec{\sigma}\: \psi)
(\varphi^t\: \sigma_2 \vec{\sigma}\: \varphi) =
-2(\psi^t\: \sigma_2\: \varphi) (\psi^t\:\sigma_2\: \varphi)$.
Thus, eq.(\ref{grav}) can be recast in the form
\begin{equation}
{\cal L_{\rm grav}} = -\frac{1}{2}\times \frac{\lambda}{M_{\rm Pl}}
({\psi}_{\alpha}\: \epsilon\: \vec{\sigma}\: 
\varphi)\: C^{\scriptscriptstyle -1}\:
({\psi}_{\beta}\: \epsilon\: \vec{\sigma} \:  \varphi) +h.c.,
\label{grav1}
\end{equation}
(suppressing all indices except those in flavor space).
Here, the reader readily recognizes the operator 
introduced in Ref.\cite{planck2}, but with a
factor $-1/2$ in front of it.
In other words, if we knew that quantum gravity 
yielded the first operator with a coefficient equal to 1, 
the second one would be generated with a 
coefficient $-1/2$, not 1.
In summary, we have to live with an uncertainty 
in the coefficient $\lambda$, and this in turn 
reflects into an uncertainty ${\cal O}(1)$ 
in our results for $U_{e3}$.

\section*{Perturbative expansion}
A natural assumption is that
the unperturbed ($0^{th}$-order) mass matrix  ${\cal M}$, 
\begin{equation}
{\cal M}=U^*\ {\rm diag}(M_i)\ U^\dagger,\mbox{ where  }
 U=\left(
\begin{array}{ccc}
U_{e1} & U_{e2} & U_{e3} \\
U_{\mu 1} & U_{\mu 2} & U_{\mu 3} \\
U_{\tau 1} & U_{\tau 2} & U_{\tau 3}, 
\end{array}
\right)
\label{zordM}
\end{equation}
(where $U_{\alpha i}$ is the usual mixing matrix,
and $M_i$ the neutrino masses) is 
generated by grand unified dynamics.\footnote{For 
instance, the complete
seesaw formula in minimal 
$SO(10)$ \cite{seesaw10} is:
${\cal M}=\frac{v^2}{V}\ (Y_\nu\, Y^{-1}\, Y_\nu^t + \xi\, Y)$
where $M_X=V$ is the vev of the singlet contained in the 126-plet, 
of the order of or somewhat smaller than the supersymmetric grand 
unification scale.
Here $Y_\nu$ is the neutrino Yukawa coupling, 
$Y$ the 126-plet coupling to the fermions and 
$\xi\; v^2/V$ is the 
effective vev of the triplet in the 126-plet
(these two give the noncanonical, or type II seesaw).}
Most of the parameters related to neutrino
oscillations are known, the major exception is given by the 
mixing element $U_{e3}$. We adopt the usual parameterization:
$|U_{e2}/U_{e1}|=\tan\theta_{12}$, $|U_{\mu2}/U_{\mu3}|=\tan\theta_{23}$
and $|U_{e3}|=\sin\theta_{13}$, or equivalently
$\theta_{12}=\omega$, $\theta_{23}=\psi$ 
and $\theta_{13}=\phi$. Note that in our approach 
$M_i$ are real and non-negative and 
we include all possible phases in the mixing matrix:
\begin{equation}
U= 
{\rm diag}(e^{if_{i}}) \ 
R(\theta_{23}) \ 
\Delta\ 
R(\theta_{13}) \ 
\Delta^*\ 
R(\theta_{12})\ 
{\rm diag}(e^{ia_{i}}) .
\label{umatr}
\end{equation}
The phase $\delta$ appearing in 
$\Delta={\rm diag}(e^{i\delta/2},1,e^{-i\delta/2})$
is the one that affects oscillations.
$a_{i}$ are the so called Majorana phases
and $f_{i}$ are usually considered as a 
part of the definition of the neutrino field. 
It is possible to rotate away the phases $f_{i}$, if the mass matrix
(\ref{zordM}) {\em is the complete mass matrix}.
However, since we are going to add another contribution to this
mass matrix, the phases $f_{i}$ of the zeroth order mass matrix 
have an impact on the complete mass matrix and thus must be retained. 
By the same token, the Majorana phases which are usually redundant
for oscillations have a dynamical role to play now.

Non-GUT effects related to a larger mass scale $M_X>M_{GUT}$
will add other contributions to the mass matrix and 
in particular will affect 
the magnitude of $U_{e3}$. Thus, let us assume 
that the mass matrix is modified as:
\begin{equation}
{\cal M}\to {\cal M}+\mu\ \lambda,
\label{eq6}
\end{equation}
with $ \mu=\frac{v^2}{M_X}$
and $\lambda$ being a matrix of dimensionless
terms as discussed in the introduction (\ref{textm}).  
The impact of the new terms on the mixing can  
be seen by forming the hermitian matrix 
$({\cal M}+\mu\lambda)^\dagger ({\cal M}+\mu\lambda)$, which is the
matrix relevant for oscillation physics.
To first order in the small parameter $\mu$, the above matrix is 
${\cal M}^\dagger {\cal M}
+\mu \lambda^{\dagger} {\cal M}
+ {\cal M}^\dagger \mu\lambda$.
Now by using eq.(\ref{zordM}) and the fact that this new mass squared
matrix must be diagonalized by a new mixing matrix resulting in corrected
eigenvalues, one can write,
\begin{equation}
U (M^2+m^\dagger M +M m) U^\dagger\equiv U' {M'}^2 {U'}^\dagger,
\mbox{ with } m=\mu\  U^t\ \lambda\ U .
\label{tech}
\end{equation}
Here $M$ and $M'$ are the diagonal matrices with 
neutrino masses at   
$0^{th}$ and at $1^{st}$ order in $\mu$.
It is clear from eq.(\ref{tech}) that the new mixing matrix  
can be written as
\begin{equation}
U'=U\ (1+i\delta\theta) ,
\label{eq8}
\end{equation}
where $\delta\theta$ is a hermitian 
matrix that appears at first order in $\mu$. 

{}From eq.(\ref{tech}) one obtains 
\begin{equation}
M^2+m^\dagger M +M m =
{M'}^2 + [ i\delta\theta, M^{'2}].
\label{tech1}
\end{equation}
Therefore to
first order in $\mu$,  
the mass squared differences $\Delta M^2_{ij}=M_i^2-M_j^2$ 
get modified as:
\begin{equation}
\Delta M^{'2}_{ij} = \Delta M^2_{ij} + 2\; (M_i \mbox{Re}[m_{ii}]
-M_j  \mbox{Re}[m_{jj}])
\label{dms}
\label{eq9}
\end{equation}
and the new contributions to the mixing matrix are:
\begin{equation}
\delta \theta_{ij} = 
  \frac{i\; \mbox{Re}[m_{ij}](M_i+M_j)}{\Delta M^{'2}_{ij}} - 
\frac{\mbox{Im}[m_{ij}](M_i-M_j)}{\Delta M^{'2}_{ij}} .
\label{interm}
\label{eq10}
\end{equation}
The diagonal elements of $\delta \theta_{ii}$ are
undetermined, as follows from the  
phase invariance of eq.(\ref{tech}). 
Thus we set them to zero.
Putting together eq.(\ref{interm}) and the 
definition of $m$ in eq.(\ref{tech}), 
we obtain the contribution to $U_{e3}$
at ${\cal O}(\mu)$: 
\begin{equation}
\delta U_{e3}=\begin{array}[t]{l}\displaystyle
{\mu(M_3+M_1) }/{\Delta M_{31}^{'2}}  \times 
U_{e1}\ \mbox{Re}(U^t \lambda U)_{13}\\[1pt]
\displaystyle -i\ {\mu(M_3-M_1)}/{\Delta M_{31}^{'2}}  \times 
U_{e1}\ \mbox{Im}(U^t \lambda U)_{13}
+ (1\to 2) .
\end{array}
\label{Main}
\label{eq11}
\end{equation}
One observes that it is the atmospheric neutrino mass difference
$\Delta M_{31}^{'2}$ that enters into the final expression for $\delta U_{e3}$.
The solar neutrino mass difference has no role in determining
the magnitude of this effect. If the corrected 
mass squared difference $\Delta M^{'2}_{31}$ is almost
the same as the original mass squared difference $\Delta M^{2}_{31}$
(as is true for Planck scale effects), then 
the above formula simplifies to the one given
below:
\begin{equation}
\delta U_{e3}=
{\mu }\ U_{e1} \left( \frac{\mbox{Re}(U^t \lambda U)_{13}}{M_3-M_1}
-i  \
\frac{\mbox{Im}(U^t \lambda U)_{13}}{M_3+M_1} \right)
+ (1\to 2) .
\label{main}
\end{equation}
Eq.(\ref{Main}) should be used to calculate the contribution 
for scales less than the Planck scale.
Eqs.(\ref{Main}) and (\ref{main}) are our main results.
They describe the contribution to $U_{e3}$ coming from
a perturbation above the GUT scale.\newline 
{}From eqs.(\ref{dms}) and (\ref{interm}) 
one can also obtain the size of the deviations 
of $\theta_{12}$ and $\theta_{23}$ from maximal values and study 
the stability of the $\Delta M^2_{ij}$ under effects due 
to physics above the grand unified scale. In this paper however,
we focus on $U_{e3}$.

\section*{Numerical results}
To estimate $\delta U_{e3}$ numerically from eq.(\ref{main}) we need
to know the 
mixing terms $U_{\alpha j}$, the mass squared 
differences $\Delta M^2_{ij}$ 
and the absolute neutrino masses $M_1$, $M_2$ and $M_3$. 
While the former two 
can be taken either from experimental data (except for the 
phases that at present remain unknown) or from 
a specified theoretical model, the `absolute' neutrino masses 
cannot be obtained from oscillation experiments. See
for instance \cite{vissa}.
We take the solar neutrino mass difference $\Delta M^2_{21} = 
7.1 \times 10^{-5}$~eV$^2$ and the solar mixing angle
$\theta_{12} = 34^\circ$.
The atmospheric neutrino mass
difference is taken to be $\Delta M^2_{31} = 2.8 \times 10^{-3} \mbox{eV}^2$ 
and the mixing angle $\theta_{23}$ to be $45^\circ$.
The usual CP violating phase can be taken as~$\delta=0$.

To get a numerical estimate for $\delta U_{e3}$  
we should consider what we know on `absolute' neutrino masses.
More precisely there are a certain number of masses:\\ 
{1)} The mass of the lightest neutrino $M$,
equal to either $M_1$ or to $M_3$ for the 
normal and the inverted hierarchy 
respectively.\\
{2)} The ``cosmological'' mass $M_{\rm cosm}=M_1+M_2+M_3$, which can be
determined from the distribution of matter on large-scales and 
the anisotropy of the cosmic microwave background 
radiation. From the first paper in \cite{WMAP}, we quote
$M_{\rm cosm}<0.71$~eV 
at 95 $\%$ CL.
This bound is valid under the simplest 
cosmological assumptions and depends on them rather crucially.
See {\em e.g.}\ \cite{disc}.\\
{3)} The effective mass of the electron neutrino coming 
from the tritium 
experiment 
$M_{\nu_e}^2=\sum |U_{ei}^2| M_i^2$ $< (2.2\mbox{ eV})^2$ 
at 95 $\%$ CL \cite{trit}.\\
{4)} Neutrinoless double beta decay 
constrains $m_{ee}=|\sum_i U_{ei}^2 M_i|
<0.38\; h$~eV at 95 $\%$ CL \cite{hm}
($h=0.6-2.8$ quantifies the 
uncertainty in nuclear matrix elements).
If the Majorana phases are the same, {\em e.g.}\ zero,
there is no cancellation of the 3 contributions and 
the bound implies
$M_1 \approx M_{\nu_{e}} \approx M_{\rm cosm}/3<0.38\; h$~eV. 
If the Majorana phases produce the largest possible
cancellation, this relaxes to $< 1.2\; h$~eV \cite{vissa}.

The introduction of these unknown (although constrained) quantities 
results in uncertainties in the values of $U_{e3}$ that we obtain
in our framework.
In particular, we see that the second term between brackets 
in eq.(\ref{main}) {\em decreases} when the scale $M$ 
increases, while the first term becomes instead {\em larger}.
In other words the contribution of new physics at the scale $M_X$ 
to $U_{e3}$ can be rather large in the case of 
degenerate neutrinos, although it is possible to diminish this effect 
by certain choices of 
the phases. It should be noted that 
the relative phases between the GUT 
contribution (phases $f_{i}$)
and the perturbative 
contribution influence the magnitude of $U_{e3}$.

\begin{figure}[t]
\centerline{\includegraphics[width=0.48\textwidth,angle=270]{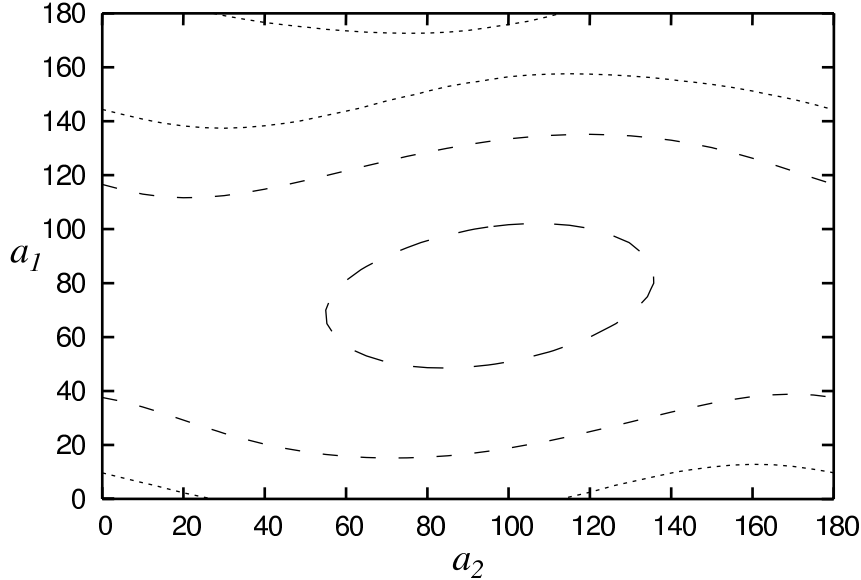}}
\caption{\em Impact of the Majorana phases on calculated
$\sin^2\! 2\theta_{13}$ values for 
Planck scale effects and a degenerate
(tritium mass) spectrum, for $a_1,a_2=0- 180^\circ$.
For illustration we show the contours where the value is: 
$1\cdot  10^{-5}$, dotted line;
$4\cdot 10^{-5}$, dashed line;
$1\cdot 10^{-4}$, long dashed line.
The value at the peak (innermost region) is $1\times 10^{-4}$. 
The value at the lower left corner (where $a_1=a_2=0$) 
is $3 \times 10^{-6}$. \label{fig1}}
\end{figure}

\feature{Planck scale effects}
When we focus on Planck scale effects, we 
assume that $\lambda_{\alpha\beta}=\lambda$ for each $\alpha$ and $\beta$.
Let us set in the unperturbed mass matrix 
the mixing $\theta_{13}=0$ and also 
$f_{i}=a_{i}=0$.  
From eq.(\ref{main}) one observes that
the surviving contribution to $U_{e3}$ is given by the first term.
All our 
results can be easily modified to include 
a specific GUT model for the unperturbed 
mass matrix ${\cal M}$, or for different 
experimental inputs and phases.
The dimensional factors 
$\mu (M_3+M_1)/\Delta M^2_{31}$ and $\mu (M_3+M_2)/\Delta M^2_{32}$
are equal with good accuracy and they are 
multiplied by $\cos\theta_{23}+\sin\theta_{23}\approx \sqrt{2}$
\cite{atm}.
Note also that 
the dependence on the solar mixing angle in this case
disappears. 
Finally  from eq.(\ref{main}) we get the range of values
for~$U_{e3}$ :
\begin{equation}
\left. U_{e3}\right|_{\rm Planck}= 
7\times 10^{-5} -  6\times 10^{-3} ,
\label{range}
\end{equation}
where the lower limit corresponds to the normal hierarchy 
for $M_1=0$ ($\Delta M^2_{32}\approx \Delta M^2_{31}=2.8
\times 10^{-3}$ eV$^2$ \cite{atm})
and the upper limit  
corresponds to the kinematical 
limit from tritium decay search, $M<2.2 \mbox{ eV}$.
This corresponds to what is usually called the ``quasi degenerate''
spectrum, where the common mass scale is much higher than the splittings
between the masses.

Some remarks are in order:\\
$(i)$~For the case of the inverted hierarchy, the numbers are practically
the same as that for the normal hierarchy.
$(ii)$~If one takes the cosmological limit on neutrino
mass, $M<0.71$ eV into account, then the upper
limit becomes $5 \times 10^{-4}$, 
which is a order of magnitude
lower than the one in eq.(\ref{range}).
$(iii)$~The range in eq.(\ref{range}) can be alternatively written as: 
\begin{equation}
\sin^2 2\theta_{13} = 2\times 10^{-8} - 1\times 10^{-4}, \mbox{ or as }
\theta_{13} = 0.004^\circ - 0.3^\circ.
\label{s2th13}
\end{equation}
$(iv)$~When the cosmological 
constraint is used, the upper 
limit of $\sin^2 2\theta_{13}$ is 
$1 \times 10^{-6}$, corresponding to 
an angle $\theta_{13}$ of $0.03^\circ$.

These numbers can be regarded as  
lower limits to $U_{e3}$ from 
Planck scale physics. They are admittedly small for a
hierarchical spectrum, but large for degenerate neutrinos.

\feature{The impact of phases}
For arbitrary values of the left and the right phases 
these numbers will change. Note that $a_3$ can always be taken
as $0$.
Putting arbitrary values of phases for the hierarchical spectrum
gives values ranging between 
$1.6-1.9 \times 10^{-8}$ for $\sin^2 2 \theta_{13}$.
So one observes that the phases have only a mild influence on the
value of $U_{e3}$.

For a degenerate spectrum,\footnote{For degenerate 
neutrinos the value of $U_{e3}$ usually scales as 
$\mu\times M$. So the results in fig.(\ref{fig1}) can be 
easily scaled to other values of $M$.  
This applies also when $\mu > v^2/M_{\rm Pl}$---see next item.} 
phases can affect the bound significantly.
For illustration in fig.(\ref{fig1}) we give $\sin^2 2 \theta_{13}$ as a
function of the right phases $a_1$ and $a_2$ as a contour plot, for
fixed arbitrary values of $f_1 = 66^\circ$, $f_2=20^\circ$ 
and $f_3= 48^\circ$. Observe that the maximum value of $1 \times 10^{-4}$
given in eq.(\ref{s2th13}) is reached for some combination of the
phases. We also see that some particular choice of phases can heavily
suppress $\sin^2 2 \theta_{13}$ and make it vanishingly small.

\begin{table}
\begin{center}
\begin{tabular}{|c|||r|r|r|||r|r|r|}
\hline
Mass spectrum &$U_{e3}$ 
& $\sin^2 2 \theta_{13}$ 
& $\theta_{13}$ & $U_{e3}$ 
& $\sin^2 2 \theta_{13}$ 
& $\theta_{13}$  \\ \hline\hline
hierarchical & $7 \cdot 10^{-4} $ & $2 \cdot 10^{-6}$ &
$0.04^\circ$ & $7 \cdot 10^{-3} $ & $2 \cdot 10^{-4}$ &
$0.4^\circ$\\ \hline
degen.\ (cosm.) & $5 \cdot 10^{-3} $ & $1 \cdot 10^{-4}$ &
$0.3 ^\circ$ & $5 \cdot 10^{-2} $ & $1 \cdot 10^{-2}$ &
$2.8 ^\circ$\\ \hline
\end{tabular}
\end{center}
\caption{\em Impact on $U_{e3}$ of a flavor blind 
mass matrix of perturbation
at the scale $M_{X}= 10^{18}$ GeV (left)
and at the scale $M_{X}= 10^{17}$ GeV (right), for the two 
types of mass spectra (first column).\label{tab1}}
\end{table}

\feature{Effect of scales below $M_{\rm Pl}$}
The lower bounds we obtained 
increase if the scale of new physics decreases
(as it is clear from eq.(\ref{main})).
In other words any model that needs 
a scale $M_X$ above $M_{GUT}$ but below the Planck mass, 
can lead to a potentially interesting contribution to the still 
unknown mixing parameter $U_{e3}$.
In this manner the scale $M_X$ can lead to  
a significant $U_{e 3}$ {\it even for non-degenerate spectra},
in contrast to the case for Planck scale effects.

We illustrate this in 
table~(\ref{tab1}) for two scales, each successively one order of
magnitude less than the Planck scale. Here we set again all the phases
to zero. We would also like to point out that for scales lower than
the Planck scale and for a degenerate neutrino spectrum, the corrections
to the solar neutrino mass difference become large (as can be seen
from eq.(\ref{dms})) 
and fine tuning of the
phases becomes unavoidable to control the corrections. This can be taken to
suggest that for scales below the Planck scale, the degenerate spectrum is
unnatural for such effects. Equivalently one can also say that if the 
degenerate spectrum is hinted at by some other phenomena, then it strongly
constrains the scale of such flavor blind contributions.

\section*{Comparison with renormalization effects}
The neutrino mass matrix 
is renormalized from the GUT scale to the electroweak scale
$M_Z\sim M_{higgs}$, 
already due to the couplings
of the standard model. This renormalization  
modifies $U_{e3}$ and 
possibly contributes to make it non-zero.
The purpose of this section is to evaluate
quantitatively this effect.

At one-loop, renormalization in the standard model is described by,
\begin{equation}
{\cal M}_{\alpha\beta}\to 
\eta_\alpha\ {\cal M}_{\alpha\beta}\ \eta_\beta, 
\ \mbox{ with }\ \eta_\alpha=\eta\cdot \exp\left[\frac{b}{(4\pi)^2}
\int^{M_{GUT}}_{M_Z}\!\!\!  y_\alpha^2(Q) \frac{dQ}{Q}\right] .
\label{wsx}
\end{equation}
The indices $\alpha$, $\beta$ should be not summed over. 
The coefficient $\eta$ (mostly due to gauge and top Yukawa couplings)
modifies the 3 masses, {\em i.e.}\ $M_i\to \eta^2 M_i$, but leaves 
the texture of the mixing matrix 
untouched since it is flavor blind.
The other contribution is different for the 3 flavors.
Thus it changes the texture and the mixing angles.
It arises since the Yukawa couplings $y_\alpha$ of the 
charged leptons are different and the numerical 
coefficient $b$ is not zero. For instance, in the 
standard model $b=1/2$, while in its supersymmetric
version, $b=-1$~\cite{babu} (here and below, we consider 
supersymmetry at the electroweak scale). 
The largest part of this correction 
is due to $\tau$ Yukawa coupling $y_\tau$. When this 
renormalization is small we can apply the perturbative
expansion
developed above. In fact, including the effect of 
$\eta$ in ${\cal M}$
(by redefining $\eta^2 {\cal M} \to {\cal M}$, so that
$\eta^2 M_i \to M_i$ and $U\to U$)
we get an expression
similar to~eq.(\ref{eq6}):
\begin{equation}
{\cal M}\to {\cal M} + \varepsilon\cdot 
{\textstyle \left(\begin{array}{ccc}
0 & 0 & {\cal M}_{e\tau} \\
0 & 0 & {\cal M}_{\mu\tau} \\
{\cal M}_{e\tau}\! & \! {\cal M}_{\mu\tau} \! & \! 2 {\cal M}_{\tau\tau} 
\end{array}\right)},
\mbox{ when }\varepsilon\equiv{\eta_\tau}/{\eta}-1 \ll 1 .
\label{ijn}
\end{equation}
In the basis where the unperturbed mass matrix 
(the first one in eq.(\ref{ijn}))
is diagonal,
the components of the term of perturbation
(the second term) are,
\begin{equation}
m_{ij}=\varepsilon\cdot [ 
M_i\ U_{\tau i}^*\ U_{\tau j} + 
U_{\tau i}\ U_{\tau j}^*\ M_j ] .
\end{equation}
The results of eqs.(\ref{tech}-\ref{main}) still 
apply, when we replace the matrix $m$ 
introduced in eq.(\ref{tech}) with the one given here
(this is the reason why we use the same symbol).

Let us focus now on the case of interest
when the mixing $U_{e3}$ is $0$ at the GUT scale. Hence we have
$U_{\tau 1}=\sin\theta_{12}\sin\theta_{23} e^{ia_1}$,
$U_{\tau 2}=-\cos\theta_{12}\sin\theta_{23} e^{ia_2}$ 
and $U_{\tau 3}=\cos\theta_{23}$ 
(from eq.(\ref{wsx}) we see that the phases $f_i$ do not 
play any role in these considerations 
and can be set to zero). The contribution
to $U_{e3}$ from renormalization is:
\begin{equation}
\delta U_{e3}=\frac{\varepsilon}{4}\cdot \sin2\theta_{12}\sin2\theta_{23} 
(f_1-f_2 )
\to \varepsilon\cdot \left( \frac{M}{0.35\ \mbox{eV}}\right)^2,
\label{edc}
\end{equation}
where $f_j=e^{i a_j} [ \cos a_j (M_3+M_j)^2 -
i \sin a_j (M_3-M_j)^2] /\Delta {M}^{'2}_{3j}$.
A crucial point to be noted is that we use this equation 
taking the values of $\theta_{12}$ and $\theta_{23}$ from  
low energy data. This is correct 
when the considered renormalization is a perturbation.
The limiting expression given above is obtained 
with two other simplifying assumptions:
Firstly we assume a degenerate mass spectrum $M_1\sim M_2\sim M$ 
at the GUT scale and
secondly the Majorana phases $a_1$ and $a_2$ are set to~$0$.

Note in passing that the contribution to
the solar splitting $\Delta M^2_{21}$ 
from $y_\tau$ renormalization is 
$2 \varepsilon \sin^2\theta_{23}$ $[(M_1^2+M_2^2)\cos2\theta_{12} 
+ \Delta M^2_{21}]$.
Thus the smallest splitting will get a contribution from the absolute neutrino 
mass $M$ \cite{lola}, unless $\theta_{12}=45^\circ$ \cite{brs}. 
This value of $\theta_{12}$ is however disfavored at $\sim 3 \sigma$ 
by combined analyses of solar and reactor neutrino data \cite{ale}.
This suggests a `naturalness' criterion, {\em viz} 
the correction should not exceed $\Delta M^2_{21}$, that means:
$\varepsilon< {\cal O}(1)\: (0.01\mbox{ eV}/M)^2$.

\begin{table}
\begin{center}
\begin{tabular}{|c|||r|r|r|||r|r|r|}
\hline
Mass spectrum &$U_{e3}$ 
& $\sin^2 2 \theta_{13}$ 
& $\theta_{13}$ & $U_{e3}$ 
& $\sin^2 2 \theta_{13}$ 
& $\theta_{13}$  \\ \hline\hline
hierarchical & $6 \cdot 10^{-6} $ & $2 \cdot 10^{-10}$ &
$0.0004^\circ$ & $1 \cdot 10^{-4} $ & $7\cdot 10^{-8}$ &
$0.01^\circ$\\ \hline
degen.\ (cosm.) & $3 \cdot 10^{-5} $ & $5 \cdot 10^{-9}$ &
$0.002^\circ$ & $7 \cdot 10^{-4} $ & $2 \cdot 10^{-6}$ &
$0.04^\circ$\\ \hline
\end{tabular}
\end{center}
\caption{\em Impact of renormalization on $U_{e3}$ 
for the two 
types of mass spectra (first column)
in the supersymmetric standard model. 
The value of $\tan\beta$ is equal to $\sqrt{3}$ in the left
part of the table and to $10$ in the right part.\label{tab2}}
\end{table}

In order to describe the impact
of renormalization on $U_{e3}$,  
the key quantity is~$\varepsilon$, given in eq.(\ref{ijn}).
To be specific, we consider the case of the supersymmetric model 
where the grand unification program is most commonly implemented
for a variety of reasons.
The tau Yukawa coupling is given by $y_\tau=m_\tau/(v \cos\beta)$
and we take the range 
$\beta=60^\circ-89^\circ$, to have perturbative 
Yukawa couplings till $M_{GUT}$
(note that this Yukawa coupling is always 
larger than the standard model one, $m_\tau/v\sim 10^{-2}$). 
A numerical integration of the one-loop 
system of equations as in \cite{nn} yields: 
\begin{equation}
-\varepsilon=7\cdot 10^{-5}- 0.15, \mbox{ when }
\beta=60^\circ- 89^\circ .
\end{equation}
The naive 
estimate $-\varepsilon\sim y_\tau^2(M_Z)\cdot \log(M_{GUT}/M_Z)/(4\pi)^2$
would give $10^{-4}- 7\cdot 10^{-2}$.

At this point we have all the ingredients to 
evaluate the contribution to $U_{e3}$ from eq.(\ref{edc}). 
The result is given in table~(\ref{tab2}) for two values of $\beta$. 
This shows that renormalization effects are not necessarily
large, even though they lead us to expect 
that {\em $U_{e3}$ is non-zero at the electroweak scale}.
Some remarks are in order:\newline
(a)~The Majorana phases affect the result. For simplicity and also
for the sake of argument, we focused the discussion 
on the case $a_1=a_2=0$.\newline
(b)~The result depends rather dramatically on the 
unknown parameter $\beta$.\newline
(c)~The naturalness criterion on the absolute mass scale 
mentioned above is satisfied for the lowest 
values of $\beta$ and also for the largest value 
of the common neutrino mass considered.\newline
(d)~The largest values of $\beta$ and of $M$
taken together, produce inaccurate estimates in perturbation
theory and should be considered only for illustration.

\section*{Discussion}
A priori, there is no strong reason to believe
that $\theta_{13}$ is exactly zero.
In the literature there are a number 
of theoretical models for $U_{e3}$.
Some of them have large values, close to the
present CHOOZ bound (for instance some minimal 
$SO(10)$ models \cite{f,m} or 
models with $U(1)$~selection rules \cite{vissa2,altr} 
where $\theta_{13}\sim \theta_C$ -- $\theta_C=13^\circ$ 
being the Cabibbo angle).
More possibilities are reviewed in~\cite{aff},
and there are also works where
$U_{e3}$ is generated at the weak scale via radiative
corrections to some texture defined at the high scale \cite{anjan}.

However, the experimental indications 
of an almost maximal mixing angle 
({\em i.e.}\ $\theta_{23}\approx 45^\circ$),
could suggest the view 
that the mixing angles take very special values
and perhaps $\theta_{13}$ is really very small.
More generally we believe that it is 
appealing to speculate on the possibility 
that $U_{e3}$ has an anomalously small value, 
say $\theta_{13}<\theta_C^2\approx 3^\circ$.
In the present paper we have argued 
that such a small value  might be 
a window for physics above the grand unification 
scale. In this sense even a {\em negative} result from 
future search of $\theta_{13}$ could be of great interest. 

Our analysis concerns the aforementioned case.
We assumed that $U_{e3}$ is zero at the GUT scale 
and we addressed the question on whether the physics above the GUT scale 
can be responsible for a non-zero value of $U_{e3}$.
The outcome is
not discouraging, especially if neutrinos are mass degenerate,
or if the neutrino masses receive contributions 
from other scales below $M_{\rm Pl}$.
In fact, a 
sizable part of the range of values of $\theta_{13}$  obtained here 
is within reach of the high performance neutrino 
factories. See for instance \cite{geer}.
The largest values
of $\theta_{13}$ can have an effect on supernova neutrino fluxes
via the MSW effect \cite{msw}. Indeed, the 'flip probability' in
the supernova mantle due to the atmospheric neutrino $\Delta M^2$ 
is approximatively given by
$ P_f\approx  \exp [- \xi\:
(U_{e3}^2/10^{-5})  ] $ where $\xi=(15\mbox{ MeV}/E_\nu)^{2/3}$. 
The tentative lower limit we discussed 
suggest that $\theta_{13}$ could well be 
at the border of the region of adiabaticity and hence can lead to
a distortion in the spectrum of supernova neutrinos \cite{ca}.


\footnotesize 
\frenchspacing
\begin{multicols}{2}

\end{multicols}
\end{document}